\documentclass{ws-procs9x6}
\usepackage{clshan-math,clshan-dm-dd}
\def\lsim{\:\raisebox{-0.5ex}{$\stackrel{\textstyle<}{\sim}$}\:}
\def\gsim{\:\raisebox{-0.5ex}{$\stackrel{\textstyle>}{\sim}$}\:}
\begin{document}
\title{HOW PRECISELY COULD WE IDENTIFY WIMPS \\
       MODEL--INDEPENDENTLY WITH \\
       DIRECT DARK MATTER DETECTION EXPERIMENTS}
\author{MANUEL DREES}
\address{Physikalisches Institut and Bethe Center of Theoretical Physics \\
         Universit\"at Bonn, D-53115 Bonn, Germany \\  
         E-mail: drees@th.physik.uni-bonn.de}
\author{CHUNG-LIN SHAN\footnote{Speaker}}
\address{School of Physics and Astronomy, Seoul Nat'l Univ.,
         Seoul 151-747, Republic of Korea \\
         E-mail: cshan@hep1.snu.ac.kr}
\begin{abstract}
  In this talk we present data analysis methods for reconstructing the
  mass and couplings of Weakly Interacting Massive Particles (WIMPs)
  by using directly future experimental data (i.e., measured recoil
  energies) from direct Dark Matter detection.  These methods are
  independent of the model of Galactic halo as well as of WIMPs.
  The basic ideas of these methods and the feasibility and
  uncertainties of applying them to direct detection experiments with
  the next generation detectors will be discussed.
\end{abstract}
\section{Introduction}
Weakly Interacting Massive Particles (WIMPs) $\chi$ arising in several
extensions of the Standard Model of electroweak interactions with
masses roughly between 10 GeV and a few TeV are one of the leading
candidates for Dark Matter\cite{{SUSYDM96}, {Bertone05}}.  Currently,
the most promising method to detect different WIMP candidates is the
direct detection of the recoil energy deposited by elastic scattering
of ambient WIMPs on the target nuclei\cite{{Smith90}, {Lewin96}}. The
differential event rate for elastic WIMP--nucleus scattering is given
by\cite{SUSYDM96}:
\beq \label{eqn:dRdQ}
   \dRdQ
 = \afrac{\rho_0 \sigma_0}{2 \mchi \mrN^2} \FQ
   \int_{\vmin}^{\vmax} \bfrac{f_1(v)}{v} dv\, .
\eeq
Here $R$ is the event rate, i.e., the number of events per unit time
and unit mass of detector material, $Q$ is the energy deposited in the
detector, $\rho_0$ is the WIMP density near the Earth, $\sigma_0$ is
the total cross section ignoring the form factor suppression, $F(Q)$
is the elastic nuclear form factor, $f_1(v)$ is the one--dimensional
velocity distribution function of incident WIMPs, $v$ is the absolute
value of the WIMP velocity in the laboratory frame.  The reduced mass
$\mrN$ is defined by $m_{\rm r,N} \equiv \mchi \mN/(\mchi+\mN)$, where
$\mchi$ is the WIMP mass and $\mN$ that of the target nucleus.
Finally, $\vmin = \alpha \sqrt{Q}$ with $\alpha \equiv \sqrt{\mN / 2
  \mrN^2}$ is the minimal incoming velocity of incident WIMPs that can
deposit the energy $Q$ in the detector, and $\vmax$ is related to the
escape velocity from our Galaxy at the position of the Solar system.

The total WIMP--nucleus cross section $\sigma_0$ in
Eq.(\ref{eqn:dRdQ}) depends on the nature of the WIMP couplings on
nucleons.  Generally speaking, one has to distinguish
spin--independent (SI) and spin--dependent (SD) couplings.  Through
e.g., squark and Higgs exchanges with quarks, Majorana WIMPs e.g.,
neutralinos in the supersymmetric models, can have a SI scalar
interaction with nuclei\cite{{SUSYDM96}, {Bertone05}}:
\beq \label{eqn:sigma0SI}
   \sigmaSI
 = A^2 \afrac{\mrN}{m_{\rm r, p}}^2 \sigmapSI\, ,
   ~~~~~~~~~~~~ %12
   \sigmapSI
 = \afrac{4}{\pi} m_{\rm r, p}^2 |f_{\rm p}|^2\, ,
\eeq
where $A$ is the atomic number of target nucleus, ${m_{\rm r, p}}$ is
the reduced mass of WIMPs and protons, and $f_{\rm p}$ is the
effective $\chi \chi {\rm p p}$ four--point coupling.  Note here that
the approximation $f_{\rm n} \simeq f_{\rm p}$ predicted in most
theoretical models has been adopted and the tiny mass difference
between a proton and a neutron has been neglected.

Meanwhile, through e.g., squark and $Z$ boson exchanges with quarks,
WIMPs can couple to the spin of the target nuclei.  The total cross
section for the spin coupling can be expressed as\cite{{SUSYDM96},
  {Bertone05}}
\cheqna
\beq \label{eqn:sigma0SD}
   \sigmaSD
 = \afrac{32}{\pi} G_F^2 \~ \mrN^2 \afrac{J+1}{J}
   \bBig{\expv{S_{\rm p}} \armp + \expv{S_{\rm n}} \armn}^2\, ,
\eeq
 and
\cheqnb
\beq \label{eqn:sigmapSD}
   \sigma_{\chi {\rm (p,n)}}^{\rm SD}
 = \afrac{24}{\pi} G_F^2 \~ m_{\rm r, p}^2 \~ |a_{\rm (p, n)}|^2\, .
\eeq
\cheqn
Here $G_F$ is the Fermi constant, $J$ is the total spin of the target
nucleus, $\expv{S_{\rm (p, n)}}$ are the expectation values of the
proton and the neutron group spins, and $a_{\rm (p, n)}$ is the
effective SD WIMP coupling to protons and neutrons.
\section{Determining the WIMP mass}
It has been found that the one--dimensional velocity distribution
function of incident WIMPs, $f_1(v)$, can be solved analytically from
Eq.(\ref{eqn:dRdQ}) directly\cite{DMDDf1v} and, consequently, its
generalized moments can be estimated by\cite{DMDDmchi}
\beqn \label{eqn:moments}
 &~& \expv{v^n}(v(\Qmin), v(\Qmax))
     \non\\
 &=& \int_{v(\Qmin)}^{v(\Qmax)} v^n f_1(v) \~ dv
     \non\\
 &=& \alpha^n
     \bfrac{2 \Qmin^{(n+1)/2} r(\Qmin) / \FQmin + (n+1) I_n(\Qmin, \Qmax)}
           {2 \Qmin^{   1 /2} r(\Qmin) / \FQmin +       I_0(\Qmin, \Qmax)}\, .
\eeqn
Here $v(Q) = \alpha \sqrt{Q}$, $Q_{\rm (min, max)}$ are the minimal
and maximal cut--off energies of the experimental data set,
respectively, $r(\Qmin) \equiv (dR/dQ)_{Q = \Qmin}$ is an estimated
value of the scattering spectrum at $Q = \Qmin$, and $I_n(\Qmin,
\Qmax)$ can be estimated through the sum:
\beq \label{eqn:In_sum}
   I_n(\Qmin, \Qmax)
 = \sum_a \frac{Q_a^{(n-1)/2}}{F^2(Q_a)}\, ,
\eeq
where the sum runs over all events in the data set between $\Qmin$ and
$\Qmax$.

By requiring that the values of a given moment of $f_1(v)$ estimated
by Eq.(\ref{eqn:moments}) from two detectors with different target
nuclei, $X$ and $Y$, agree, a general expression for determining
$\mchi$ appearing in the prefactor $\alpha^n$ on the right--hand side
of Eq.(\ref{eqn:moments}) has been found as\cite{DMDDmchi-SUSY07}:
\beq \label{eqn:mchi_Rn}
   \left. \mchi \right|_{\Expv{v^n}}
 = \frac{\sqrt{\mX \mY} - \mX (\calR_{n, X} / \calR_{n, Y})}
        {\calR_{n, X} / \calR_{n, Y} - \sqrt{\mX / \mY}}\, ,
\eeq
 where
\beq \label{eqn:RnX_min}
        \calR_{n, X}
 \equiv \bfrac{2 \QminX^{(n+1)/2} r_X(\QminX) / \FQminX + (n+1) \InX}
              {2 \QminX^{   1 /2} r_X(\QminX) / \FQminX +       \IzX}^{1/n}\, ,
\eeq
and $\calR_{n, Y}$ can be defined analogously.  Here $n \ne 0$,
$m_{(X, Y)}$ and $F_{(X, Y)}(Q)$ are the masses and the form factors
of the nucleus $X$ and $Y$, respectively.  Note that, since the
general moments of $f_1(v)$ estimated by Eq.(\ref{eqn:moments}) are
independent of the WIMP--nucleus cross section $\sigma_0$, the
estimator (\ref{eqn:mchi_Rn}) of $\mchi$ can be used either for SI or
for SD scattering.

Additionally, since in most theoretical models the SI WIMP--nucleus
cross section given in Eq.(\ref{eqn:sigma0SI})
dominates\cite{{SUSYDM96}, {Bertone05}}, and on the right--hand side
of Eq.(\ref{eqn:dRdQ}) is in fact the minus--first moment of $f_1(v)$,
which can be estimated by Eq.(\ref{eqn:moments}) with $n = -1$, one
can find that\cite{DMDDmchi}
\beq \label{eqn:rho0_fp2}
   \rho_0 |f_{\rm p}|^2
 = \frac{\pi}{4 \sqrt{2}} \afrac{\mchi + \mN}{\calE A^2 \sqrt{\mN}}
   \bbrac{\frac{2 \Qmin^{1/2} r(\Qmin)}{\FQmin} + I_0}\, .
\eeq
Here $\calE$ is the exposure of the experiment which relates the
actual counting rate to the normalized rate in Eq.(\ref{eqn:dRdQ}).
Since the unknown factor $\rho_0 |f_{\rm p}|^2$ on the left--hand side
above is identical for different targets, it leads to a second
expression for determining $\mchi$\cite{DMDDmchi}:
\beq \label{eqn:mchi_Rsigma}
   \left. \mchi \right|_\sigma
 = \frac{\abrac{\mX / \mY}^{5/2} \mY - \mX (\calR_{\sigma, X} / \calR_{\sigma, Y})}
        {\calR_{\sigma, X} / \calR_{\sigma, Y} - \abrac{\mX / \mY}^{5/2}}\, .
\eeq
 Here $m_{(X, Y)} \propto A_{(X, Y)}$ has been assumed,
\beq \label{eqn:RsigmaX_min}
        \calR_{\sigma, X}
 \equiv \frac{1}{\calE_X} \bbrac{\frac{2 \QminX^{1/2} r_X(\QminX)}{\FQminX} + \IzX}\, ,
\eeq
and similarly for $\calR_{\sigma, Y}$.

In order to yield the best--fit WIMP mass as well as its statistical
error by combining the estimators for different $n$ in
Eq.(\ref{eqn:mchi_Rn}) with each other and with the estimator in
Eq.(\ref{eqn:mchi_Rsigma}), a $\chi^2$ function has been
introduced\cite{DMDDmchi}
\beq \label{eqn:chi2} 
   \chi^2(\mchi)
 = \sum_{i, j}
   \abrac{f_{i, X} - f_{i, Y}} {\cal C}^{-1}_{ij} \abrac{f_{j, X} - f_{j, Y}}\, ,
\eeq
where
\cheqna
\beq \label{eqn:fiXa}
        f_{i, X}
 \equiv \afrac{\alpha_X {\cal R}_{i, X}}{300~{\rm km/s}}^{i}\, ,
        ~~~~~~ %6
        {\rm for}~i = -1,~1,~2,~\dots,~n_{\rm max},
\eeq
 and
\cheqnb
\beq \label{eqn:fiXb}
        f_{n_{\rm max}+1, X}
 \equiv \frac{A_X^2}{\calR_{\sigma, X}} \afrac{\sqrt{\mX}}{\mchi + \mX}\~;
\eeq
\cheqn
the other $n_{\rm max} + 2$ functions $f_{i,Y}$ can be defined
analogously.  Here $n_{\rm max}$ determines the highest moment of
$f_1(v)$ that is included in the fit.  The $f_i$ are normalized such
that they are dimensionless and very roughly of order unity.  Note
that the first $n_{\rm max} + 1$ fit functions depend on $\mchi$
through the overall factor $\alpha$ and that $\mchi$ in
Eqs.(\ref{eqn:fiXa}) and (\ref{eqn:fiXb}) is now a fit parameter,
which may differ from the true value of the WIMP mass. Moreover,
$\cal C$ is the total covariance matrix.  Since the $X$ and $Y$
quantities are statistically completely independent, $\cal C$ can be
written as a sum of two terms:
\beq \label{eqn:Cij}
   {\cal C}_{ij}
 = {\rm cov}\abrac{f_{i, X}, f_{j, X}} + {\rm cov}\abrac{f_{i, Y}, f_{j, Y}}\, .
\eeq

Finally, since we require that, from two experiments with different
target nuclei, the values of a given moment of the WIMP velocity
distribution estimated by Eq.(\ref{eqn:moments}) should agree, this
means that the upper cuts on $f_1(v)$ in two data sets should be
(approximately) equal\footnote{Here the threshold energies have been
  assumed to be negligibly small.}. This requires that\cite{DMDDmchi}
\beq \label{eqn:match} 
   Q_{{\rm max}, Y}
 = \afrac{\alpha_X}{\alpha_Y}^2 Q_{{\rm max}, X}\, .
\eeq
Note that $\alpha$ is a function of the true WIMP mass.  Thus this
relation for matching optimal cut--off energies can be used only if
$\mchi$ is already known.  One possibility to overcome this problem is
to fix the cut--off energy of the experiment with the heavier target,
minimize the $\chi^2(\mchi)$ function defined in Eq.(\ref{eqn:chi2}),
and estimate the cut--off energy for the lighter nucleus by
Eq.(\ref{eqn:match}) algorithmically\cite{DMDDmchi}.

As demonstration we show some numerical results for the reconstructed
WIMP mass based on Monte Carlo simulations. The upper and lower bounds
on the reconstructed WIMP mass are estimated from the requirement that
$\chi^2$ exceeds its minimum by 1. $\rmXA{Si}{28}$ and $\rmXA{Ge}{76}$
have been chosen as two target nuclei. The scattering cross section
has been assumed to be dominated by spin--independent interactions.
The theoretically predicted recoil spectrum for the shifted Maxwellian
velocity distribution ($v_0 = 220$ km/s, $v_e = 231$ km/s)%
\cite{{SUSYDM96}, {Bertone05}, {DMDDf1v}} with the Woods-Saxon elastic
form factor\cite{{Engel91}, {SUSYDM96}, {Bertone05}} have been used.
The threshold energies of two experiments have been assumed to be
negligible and the maximal cut--off energies are set as 100 keV.  2
$\times$ 5,000 experiments with 50 events on average before cuts from
each experiment have been simulated.  In order to avoid large
contributions from very few events in the high energy range to the
higher moments\cite{DMDDf1v}, only the moments up to $n_{\rm max} = 2$
have been included in the $\chi^2$ fit.

\begin{figure}[t!]
\begin{center}
\vspace*{-1cm}
\rotatebox{-90}{\includegraphics[width=0.65\textwidth]{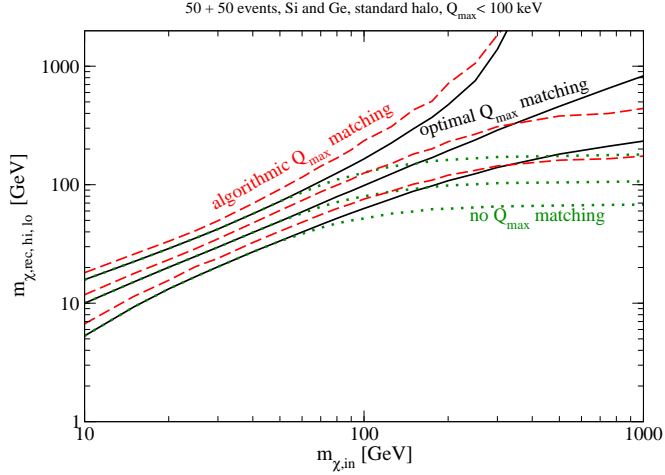}}
\vspace{-0.6cm}
\end{center}
\caption{
 Results for the median reconstructed WIMP mass
 as well as its $1 \sigma$ statistical error interval
 based on the $\chi^2$--fit in Eq.(\ref{eqn:chi2}).
 See the text for further details.
}
\label{fig:mchi_rec}
\end{figure}
In Fig.~\ref{fig:mchi_rec} the dotted (green) curves show the median
reconstructed WIMP mass and its $1 \sigma$ upper and lower bounds for
the case that both $\Qmax$ have been fixed to 100 keV.  This causes a
systematic {\em underestimate} of the reconstructed WIMP mass for
input WIMP masses $\gsim 100$ GeV\cite{DMDDmchi-SUSY07}.  The solid
(black) curves have been obtained by using Eq.(\ref{eqn:match}) for
matching the cut--off energy $Q_{\rm max, Si}$ perfectly with $Q_{\rm
  max, Ge} = 100$ keV and the true (input) WIMP mass, whereas the
dashed (red) curves show the case that $Q_{\rm max, Ge} = 100$ keV,
and $Q_{\rm max, Si}$ has been determined by minimizing
$\chi^2(\mchi)$.  As shown here, with only 50 events from one
experiment, the algorithmic process seems already to work pretty well
for WIMP masses up to $\sim 500$ GeV.  Though for $\mchi \lsim 100$
GeV $\mchi$ determined in this way {\em overestimates} its true value
by 15 to 20\%, the true WIMP mass always lies within the median limits
of the $1 \sigma$ statistical error interval up to even $\mchi = 1$
TeV\cite{DMDDmchi}.

On the other hand, in order to study the statistical fluctuation of
the reconstructed WIMP mass by algorithmic $\Qmax$ matching in the
simulated experiments, an estimator $\delta m$ has been introduced
as\cite{DMDDmchi}
\beq \label{eqn:deltam}
\renewcommand{\arraystretch}{0.5}
   \delta m
 = \left\{
    \begin{array}{l c l}
     \D   1
        + \frac{m_{\chi, {\rm lo1}} - m_{\chi, {\rm in }}}
               {m_{\chi, {\rm lo1}} - m_{\chi, {\rm lo2}}}\, , & ~~~~~~ & %6
        {\rm if}~m_{\chi, {\rm in }} \leq m_{\chi, {\rm lo1}}\~;
     \\ & & \\
    \D   \frac {m_{\chi, {\rm rec}} - m_{\chi, {\rm in }}}
               {m_{\chi, {\rm rec}} - m_{\chi, {\rm lo1}}}\, , &        &
        {\rm if}~m_{\chi,{\rm lo1}} < m_{\chi, {\rm in }} < m_{\chi, {\rm rec}}\~;
    \\ & & \\
    \D   \frac {m_{\chi, {\rm rec}} - m_{\chi, {\rm in }}}
               {m_{\chi, {\rm hi1}} - m_{\chi, {\rm rec}}}\, , &        &
        {\rm if}~m_{\chi,{\rm rec}} < m_{\chi, {\rm in }} < m_{\chi, {\rm hi1}}\~;
    \\ & & \\
    \D   \frac {m_{\chi, {\rm hi1}} - m_{\chi, {\rm in }}}
               {m_{\chi, {\rm hi2}} - m_{\chi, {\rm hi1}}} - 1 \, , &   &
        {\rm if}~m_{\chi, {\rm in }} \geq m_{\chi,{\rm hi1}}\, .
    \end{array}
   \right.
\eeq
Here $m_{\chi, {\rm in }}$ is the true (input) WIMP mass, $m_{\chi,
  {\rm rec}}$ its reconstructed value, $m_{\chi, {\rm lo1(2)}}$ are
the $1 \~ (2) \~ \sigma$ lower bounds satisfying
\mbox{$\chi^2(m_{\chi, {\rm lo(1, 2)}}) = \chi^2(m_{\chi, {\rm rec}})
  + 1 \~ (4)$}, and $m_{\chi, {\rm hi1(2)}}$ are the corresponding $1
\~ (2) \~ \sigma$ upper bounds.

\begin{figure}[t]
\begin{center}
\rotatebox{-90}{\includegraphics[width=0.45\textwidth]{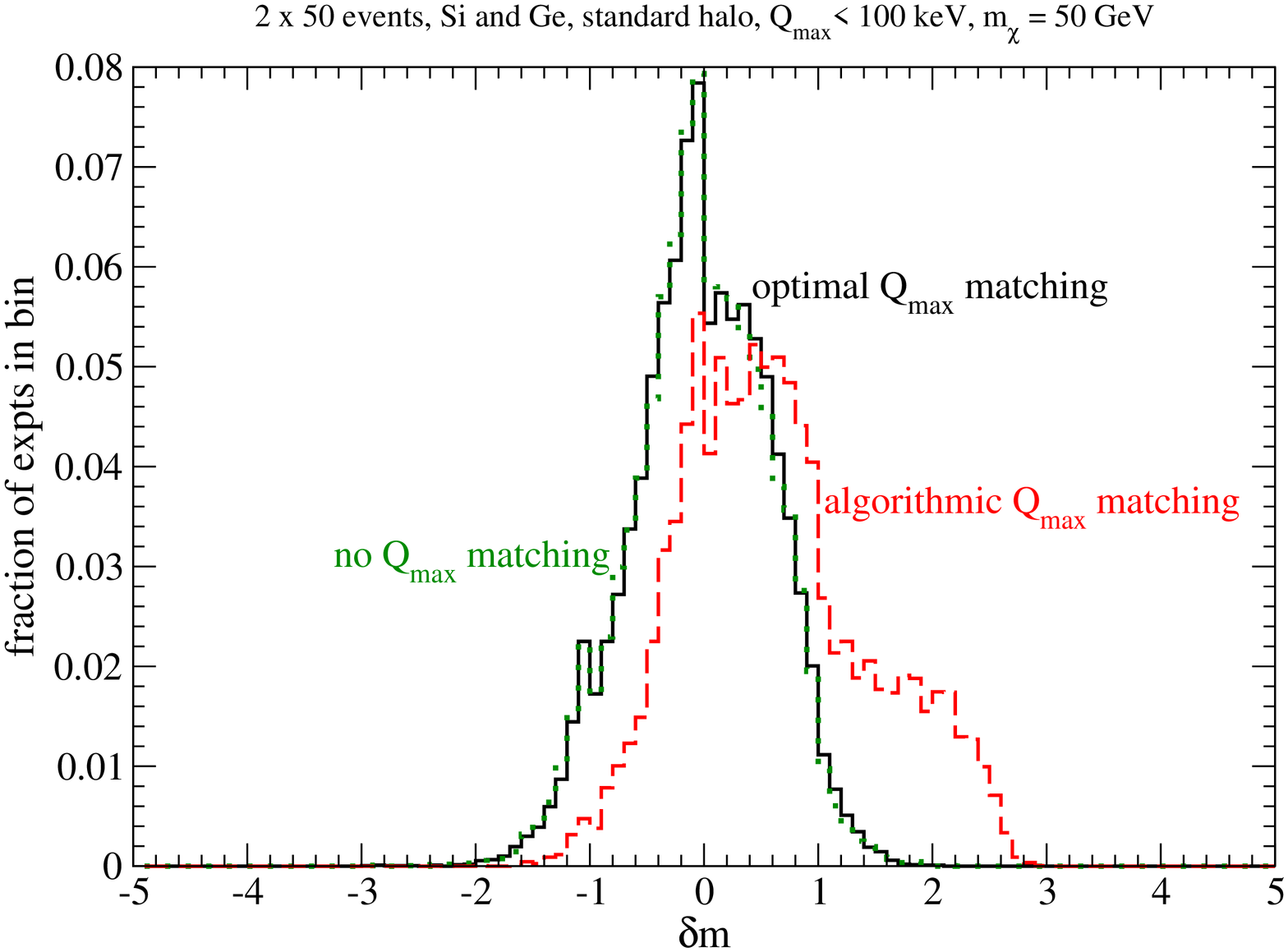}}  \hspace{-0.025\textwidth}
\rotatebox{-90}{\includegraphics[width=0.45\textwidth]{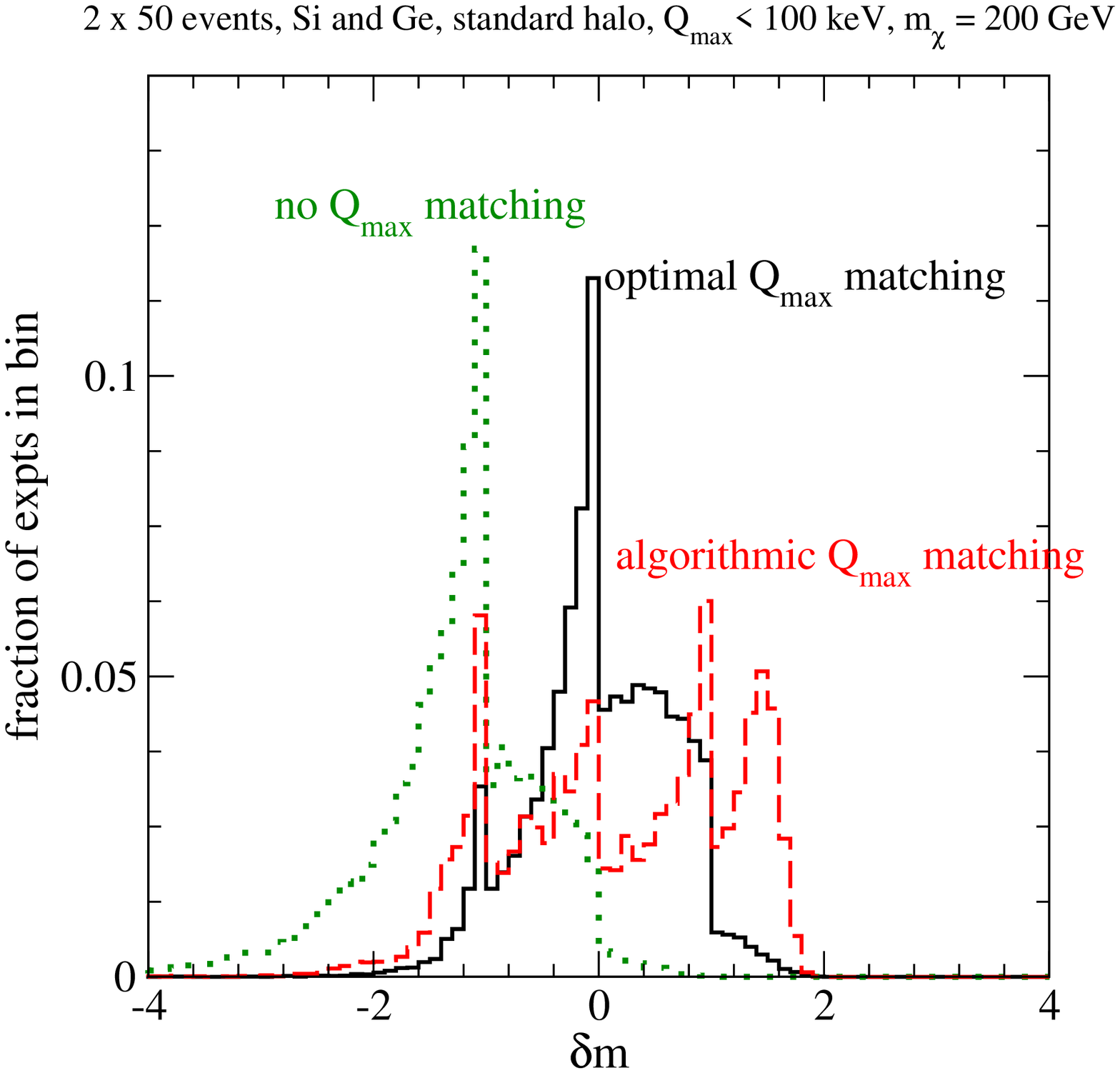}} \\
\vspace*{-0.5cm}
\end{center}
\caption{
 Normalized distribution of the estimator $\delta m$
 defined in Eq.(\ref{eqn:deltam})
 for WIMP masses of 50 GeV (left) and 200 GeV (right).
 Parameters and notations are as in Fig.~\ref{fig:mchi_rec}.
 Here the bins at $\delta m = \pm 5$ are overflow bins,
 i.e., they also contain all experiments with $|\delta m| \ge 5$.
}
\label{fig:del50}
\end{figure}
Figures \ref{fig:del50} show the distribution of the estimator $\delta
m$ calculated from 5,000 simulated experiments for WIMP masses of 50
GeV (left) and 200 GeV (right).  For the lighter WIMP mass, simply
fixing both $\Qmax$ values to 100 keV still works fine (the dotted
(green) curves in Fig.~\ref{fig:mchi_rec}).  However, the
distributions for both fixed $\Qmax$ and optimal $\Qmax$ matching show
already an asymmetry of the statistical uncertainties with $ m_{\chi,
  {\rm hi1}} - m_{\chi, {\rm rec}} > m_{\chi, {\rm rec}} - m_{\chi,
  {\rm lo1}}$.  The overestimate of light WIMP masses reconstructed by
algorithmic $\Qmax$ matching shown in Fig.~\ref{fig:mchi_rec} is also
reflected by the dashed (red) histogram here, which has significantly
more entries at positive values than at negative values.  Moreover,
these distributions also indicate that the statistical uncertainties
estimated by minimizing $\chi^2(\mchi)$ are in fact overestimated,
since nearly 90\% of the simulated experiments have $|\delta m| \leq
1$\cite{DMDDmchi}, much more than $\sim$ 68\% of the experiments, which
a usual $1 \sigma$ error interval should contain.

For the heavier WIMP mass of 200 GeV, as shown in the right frame of
Figs.~\ref{fig:del50}, the situation becomes less favorable.  While
the distributions for both fixed $\Qmax$ and optimal $\Qmax$ matching
look more non--Gaussian but more concentrated on the median values,
the distribution for algorithmic $\Qmax$ matching spreads out in the
range $-1 < \delta m < 2$.  It has even been observed that, for larger
samples (e.g., with 500 events on average) the outspread distribution
becomes broader\cite{DMDDmchi}.  Hence, the statistical fluctuation by
the algorithmic procedure for determining $\Qmax$ of the experiment
with the lighter target nucleus by minimizing $\chi^2$ could be
problematic for the determination of $\mchi$ if WIMPs are heavy.

\section{Estimating the SI WIMP--proton coupling}
As shown in the previous section, by combining two experimental data
sets, one can estimate the WIMP mass $\mchi$ without knowing the
WIMP--nucleus cross section $\sigma_0$.  Conversely, by using
Eq.(\ref{eqn:rho0_fp2}), one can also estimate the SI WIMP--proton
coupling, $|f_{\rm p}|^2$, from experimental data directly {\em
  without} knowing the WIMP mass\cite{DMDDfp2-IDM2008}.

In Eq.(\ref{eqn:rho0_fp2}) the WIMP mass $\mchi$ on the right--hand
side can be determined by the method described in Sec.~2, $r(\Qmin)$
and $I_0$ can also be estimated from one of the two data sets used for
determining $\mchi$ or from a third experiment. However, due to the
degeneracy between the local WIMP density $\rho_0$ and the coupling
$|f_{\rm p}|^2$, one {\em cannot} estimate each one of them without
making some assumptions. The simplest way is making an assumption for
the local WIMP density $\rho_0$.

\begin{figure}[b!]
\begin{center}
\includegraphics[width=0.5\textwidth]{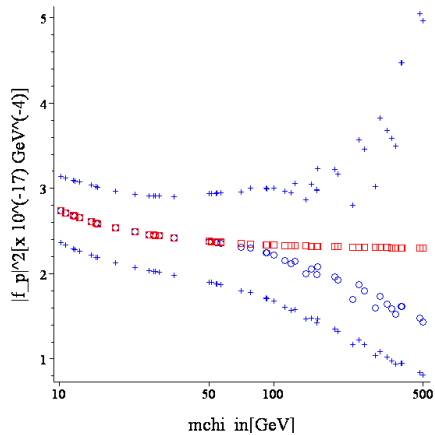}
\vspace*{-0.25cm}
\end{center}
\caption{
 The reconstructed SI WIMP--proton coupling
 as a function of the input WIMP mass.
 The (red) squares indicate
 the input WIMP masses and
 the true values of the coupling.
 The (blue) circles and the (blue) crosses indicate
 the reconstructed couplings and
 their $1 \sigma$ statistical errors.
 Parameters are as in Fig.~\ref{fig:mchi_rec},
 in addition $\sigmapSI$ has been set as $10^{-8}$ pb.
 See the text for further details.
}
\label{fig:fp2_rec}
\end{figure}

Figure \ref{fig:fp2_rec} shows the reconstructed SI WIMP--proton
coupling as a function of the input WIMP mass..  The WIMP mass has
again been reconstructed with $\rmXA{Si}{28}$ and $\rmXA{Ge}{76}$.  In
order to avoid complicated calculations of the correlation between the
error on the reconstructed $\mchi$ and that on the estimator of $I_0$,
a second, independent data set with Ge has been chosen as the third
target for estimating $I_0$.  Parameters are as in
Fig.~\ref{fig:mchi_rec}, except that the SI WIMP--proton cross section
has been set as $10^{-8}$ pb.

It can be seen that the reconstructed $|f_{\rm p}|^2$ are {\em
  underestimated} for WIMP masses $\gsim~100$ GeV.  This systematic
deviation is caused mainly by the underestimate of $I_0$.  However, in
spite of this systematic deviation the true value of $|f_{\rm p}|^2$
always lies within the $1 \sigma$ statistical error interval.
Moreover, for a WIMP mass of 100 GeV, one could in principle already
estimate the SI WIMP--proton coupling with a statistical uncertainty
of only $\sim$ 15\% with just 50 events from each experiment. Note
that this is much smaller than the systematic uncertainty of the local
Dark Matter density (of a factor of 2 or even larger).

\section{Determining ratios of WIMP--nucleon cross sections}
So far we have discussed only the case that the spin--independent
WIMP--nucleus interaction dominates.  In this section we turn to
consider the case of the spin--dependent cross section as well as of a
general combination of these two cross sections.

\subsection{Determining the \boldmath$\armn / \armp$ ratio}
Consider at first the case that the SD WIMP--nucleus interaction
dominates.  By substituting $\sigmaSD$ in Eq.(\ref{eqn:sigma0SD}) and
$\expv{v^{-1}}$ estimated by Eq.(\ref{eqn:moments}) into
Eq.(\ref{eqn:dRdQ}) and combining two data sets with different target
nuclei, an expression for the ratio between two SD WIMP-nucleon
couplings can be given as
\beq \label{eqn:ranapSD}
   \afrac{\armn}{\armp}_{\pm, n}^{\rm SD}
% =-\frac{\SpX \pm \SpY \abrac{\calR_{J, n, X} / \calR_{J, n, Y}} }
%        {\SnX \pm \SnY \abrac{\calR_{J, n, X} / \calR_{J, n, Y}} }
 =-\frac{\SpX \calR_{J, n, Y} \pm \SpY \calR_{J, n, X}}
        {\SnX \calR_{J, n, Y} \pm \SnY \calR_{J, n, X}}\, ,
\eeq
 with
\beq \label{eqn:JnX}
        \calR_{J, n, X}
 \equiv \bbrac{\Afrac{J_X}{J_X + 1}
               \frac{\calR_{\sigma, X}}{\calR_{n, X}}}^{1/2}\, ,
\eeq
and similarly for $\calR_{J, n, Y}$, where $n \ne 0$. Note that $\armn
/ \armp$ can be estimated from experimental data directly through
estimating $\calR_{n, X}$, $\calR_{\sigma, X}$ and two $Y$ terms by
Eqs.(\ref{eqn:RnX_min}) and (\ref{eqn:RsigmaX_min})\footnote{Note that
  the form factor $\FQ$ here must be chosen for the SD cross section.}
{\em without} knowing the WIMP mass.

\begin{figure}[b!]
\includegraphics[width=0.45\textwidth]{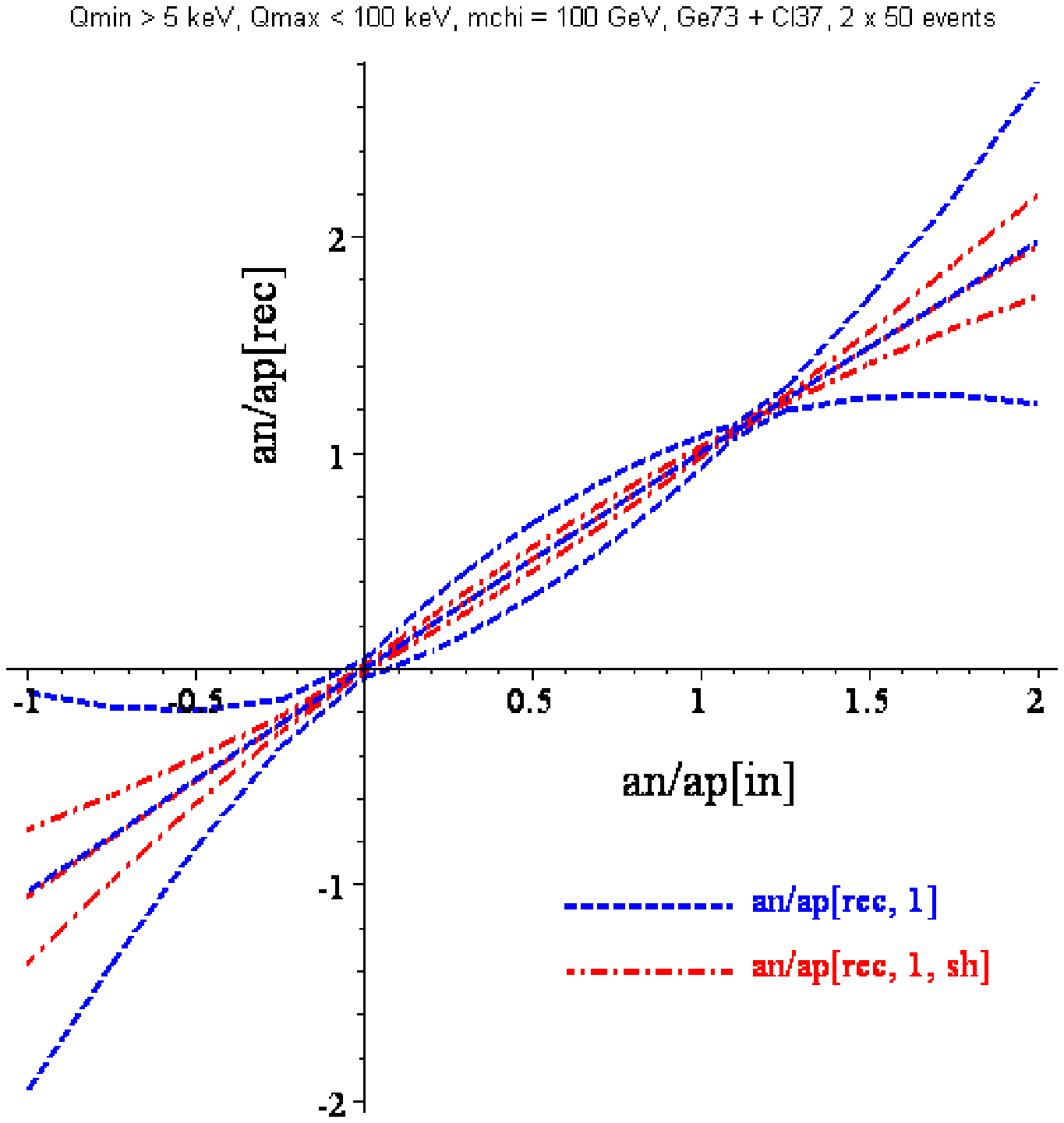} \hspace{0.025\textwidth}
\includegraphics[width=0.45\textwidth]{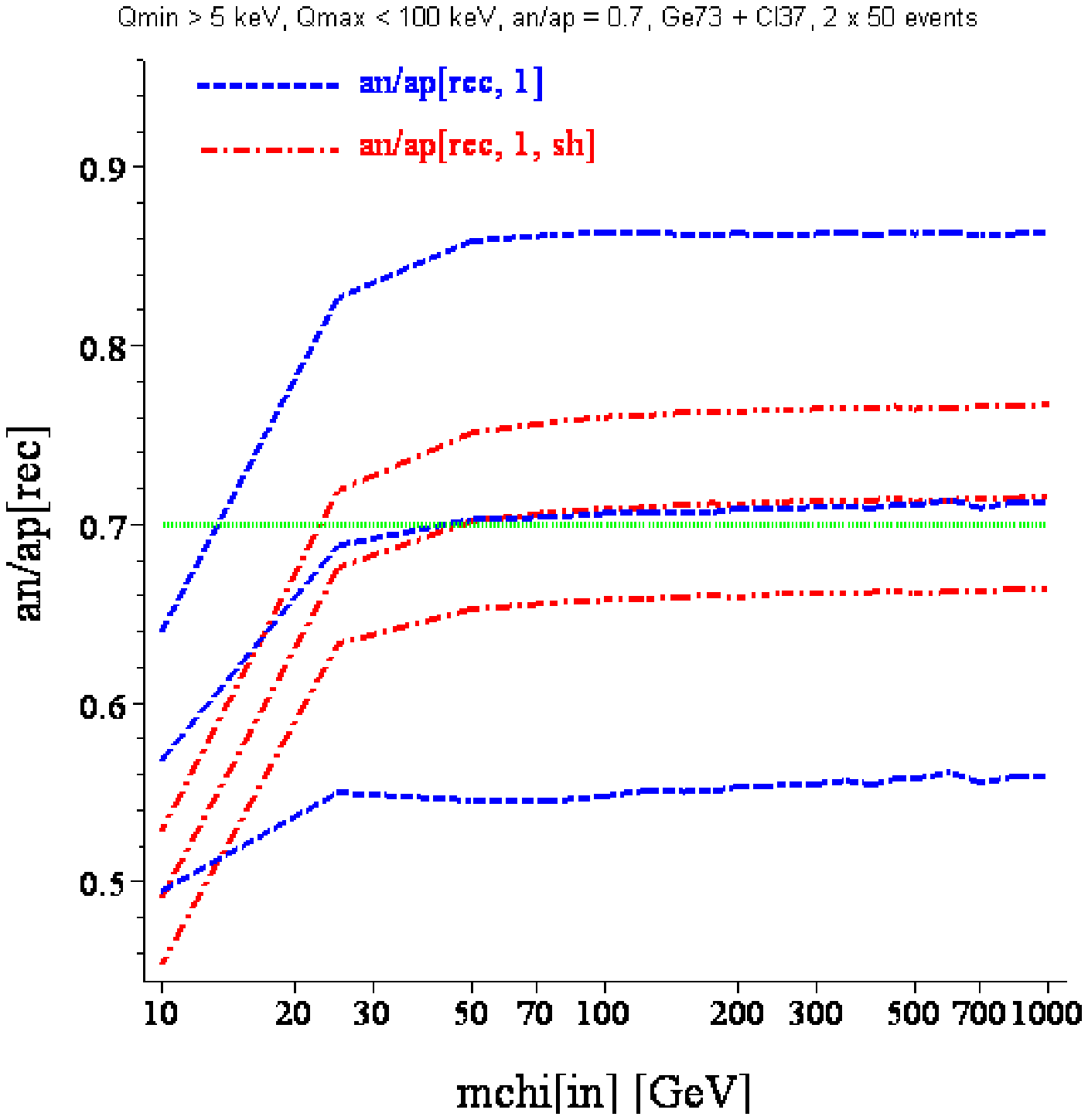}
\caption{ Preliminary results for the reconstructed $(\armn /
  \armp)^{\rm SD}$ estimated by Eq.(\ref{eqn:ranapSD}) with $n = 1$
  as functions of the true (input) $\armn / \armp$ (left frame, for a
  WIMP mass of 100 GeV) and as functions of the input WIMP mass
  $\mchi$ (right frame, for $\armn / \armp = 0.7$), respectively.  See
  the text for further details.  }
\label{fig:ranap_SD}
\end{figure}

Because the couplings in Eq.(\ref{eqn:sigma0SD}) are squared, we have
two solutions for $\armn / \armp$ here; if exact ``theory'' values for
${\cal R}_{J, n, (X, Y)}$ are taken, these solutions coincide for
$\armn /\armp = -\SpX / \SnX$ and $-\SpY / \SnY$, which depends only on the
properties of target nuclei\footnote{Some relevant spin values of the
  nuclei used for our simulations shown in this paper are given in
  Table.~1.}. Moreover, one of these two solutions has a pole at the
middle of two intersections, which depends simply on the signs of
$\SnX$ and $\SnY$: since $\calR_{J, n, X}$ and $\calR_{J, n, Y}$ are
always positive, if both of $\SnX$ and $\SnY$ are positive or
negative, the ``minus'' solution $(\armn / \armp)^{\rm SD}_{-, n}$
will diverge and the ``plus'' solution $(\armn / \armp)^{\rm SD}_{+,
  n}$ will be the ``inside'' solution, which has a smaller statistical
uncertainty (see Figs.~\ref{fig:ranap_SD}); in contrast, if the signs
of $\SnX$ and $\SnY$ are opposite, the ``minus'' solution $(\armn /
\armp)^{\rm SD}_{-, n}$ will be the ``inside'' solution.

\begin{table}[t!]
\tbl{
 List of the relevant spin values of the nuclei
 used for simulations shown in this paper
 (Data from Ref.~10).
}
{
\renewcommand{\arraystretch}{1.3}
%\begin{center}
\begin{tabular}{c  c  c  c  c  c}
\hline
 \makebox[1.5cm][c]{nucleus}        &
 \makebox[1  cm][c]{$Z$}            & \makebox[1.2cm][c]{$J$}     &
 \makebox[1.5cm][c]{$\Srmp$}        & \makebox[1.5cm][c]{$\Srmn$} &
 \makebox[1.8cm][c]{$-\Srmp/\Srmn$} \\
\hline
 $\rmXA{O}{17}$  &  8 & 5/2 &                   0     & 0.495 &     0     \\
\hline
 $\rmXA{Na}{23}$ & 11 & 3/2 &                   0.248 & 0.020 & $-$12.40  \\
\hline
 $\rmXA{Cl}{37}$ & 17 & 3/2 & \hspace{-1.8ex}$-$0.058 & 0.050 &     1.16  \\
\hline
 $\rmXA{Ge}{73}$ & 32 & 9/2 &                   0.030 & 0.378 &  $-$0.079 \\
\hline
\end{tabular}
%\caption{
% List of the relevant spin values of the nuclei
% used for simulations shown in this paper
% (Data from Ref.~\cite{Giuliani05}).
\label{tab:J_Sp_Sn}
%\end{center}
}
\end{table}
Figures \ref{fig:ranap_SD} show the reconstructed $(\armn /
\armp)^{\rm SD}$ estimated by Eq.(\ref{eqn:ranapSD}) with $n = 1$ as
functions of the true (input) $\armn / \armp$ for a WIMP mass of 100
GeV (left) and as functions of the input WIMP masses for $\armn /
\armp = 0.7$ (right), respectively.  The shifted Maxwellian velocity
distribution with a form factor calculated in the thin-shell
approximation for the SD cross section\cite{{Lewin96}, {Klapdor05}}
has been used.  Parameters are as earlier, except that the minimal
cut--off energy has been increased to 5 keV for both experiments.
Here we have chosen $\rmXA{Ge}{73}$ and $\rmXA{Cl}{37}$ as two target
nuclei in order to test the range of interest $0 \le \armn/\armp \le
1$\cite{{Bednyakov04}, {Ellis08}}; and $(\armn / \armp)^{\rm SD}_{+/-,
  n}$ are thus the inside/outside solutions.

For estimating the statistical error on $\armn / \armp$, one needs to
estimate the counting rate at the threshold energy, $r(\Qmin)$, and
its statistical error, $\sigma(r(\Qmin))$.  It has been found that,
instead of $\Qmin$, one can estimate the counting rate and its
statistical error at the shifted point $Q_{s, 1}$ (from the central
point of the first bin, $Q_1$)%
\cite{DMDDf1v}:
\beq \label{eqn:Qsn}
   Q_{s, 1}
 = Q_1 + \frac{1}{k_1} \ln\bfrac{\sinh(k_1 b_1 / 2)}{k_1 b_1 / 2}\, ,
\eeq
where $k_1$ is the logarithmic slope of the reconstructed recoil
spectrum in the first $Q$--bin and $b_1$ is the bin width.  We see in
the right frame of Figs.~\ref{fig:ranap_SD} very clearly that, for
WIMP masses $\gsim 30~{\rm GeV}$, the $1 \sigma$ statistical error
estimated with $Q_{s, 1}$ (the dash--dotted (red) lines, labeled with
``sh'') is $\sim$ 7\%, only 1/3 of the error estimated with $\Qmin$
(the dashed (blue) lines).

One more advantage with using $Q_{s, 1}$ instead of $\Qmin$ is that
the statistical error on $\armn / \armp$ estimated with different $n$
(namely with different moments of the WIMP velocity distribution) at
$Q = Q_{s, 1}$ are almost equal.  Therefore, since
\beq \label{eqn:JmaX}
   \calR_{J, -1, X}
 = \bbrac{\afrac{J_X}{J_X + 1}
          \frac{2 \~ r_X(Q_{X, s, 1})}{\calE_X F_X^2(Q_{X, s, 1})}}^{1/2}\, ,
\eeq
one needs thus only events in the low energy range ($\sim$ 20 events
between 5 and 15 keV in our simulations) for estimating $\armn /
\armp$.
\subsection{Determining the \boldmath$\sigma_{\chi \rm p/n}^{\rm SD} /
  \sigmapSI$ ratios} 
Now let us combine WIMP--nucleus scattering induced by both SI and SD
interactions given in Eqs.(\ref{eqn:sigma0SI}) and
(\ref{eqn:sigma0SD}) (with the corresponding form factors).  By
modifying $\FQ$ and $I_n$ in the estimator (\ref{eqn:moments}) of the
moments of the WIMP velocity distribution, the ratio of the SD
WIMP-proton cross section to the SI one can be solved analytically as%
\footnote{In this section we consider only the case of $\sigmapSD$,
  but all formulae given here can be applied straightforwardly to the
  case of $\sigmanSD$ by exchanging n $\getsto$ p.}$^{,}$
\footnote{$\Qmin$ appearing in this section can be replaced by $Q_{s,
    1}$ everywhere.}
\beq
   \frac{ \label{eqn:rsigmaSDpSI}\sigmapSD}{\sigmapSI}
% = \frac{\FSIQminY (\calR_{m, X}/\calR_{m, Y}) - \FSIQminX}
%        {\calCpX \FSDQminX - \calCpY \FSDQminY (\calR_{m, X} / \calR_{m, Y})}
 =-\frac{\FSIQminX \calR_{m, Y} - \FSIQminY \calR_{m, X}}
        {\calCpX \FSDQminX \calR_{m, Y} - \calCpY \FSDQminY \calR_{m, X}}\, .
\eeq
Here
\beq \label{eqn:RmX}
        \calR_{m, X}
 \equiv \frac{r_X(\QminX)}{\calE_X \mX^2}\, ,
\eeq
and
\beq \label{eqn:CpX}
        \calCpX
 \equiv \frac{4}{3} \afrac{J_X + 1}{J_X} \bfrac{\SpX + \SnX (\armn/\armp)}{A_X}^2\~;
\eeq
$\calR_{m, Y}$ and $\calCpY$ can be defined analogously.  Note that a
``minus ($-$)'' sign appears in the expression (\ref{eqn:rsigmaSDpSI}).

By introducing a third target having {\em only} the SI interaction
with WIMPs, $\armn / \armp$ appearing in $\calCpX$ and $\calCpY$ can
again be solved analytically as
\beq \label{eqn:ranapSISD}
   \afrac{\armn}{\armp}_{\pm}^{\rm SI+SD}
 = \frac{-\abrac{\cpX \snpX - \cpY \snpY} \pm \sqrt{\cpX \cpY} \vbrac{\snpX - \snpY}}
        {\cpX \snpX^2 - \cpY \snpY^2}\, .
\eeq
 Here
\beqn \label{eqn:cpX}
          \cpX
 &\equiv& \frac{4}{3} \afrac{J_X+1}{J_X} \afrac{\SpX}{A_X}^2
          \FSDQminX
          \non\\
 &~&      ~~~~~~~~ \times %8
          \bbrac{\FSIQminZ \afrac{\calR_{m,Y}}{\calR_{m,Z}} - \FSIQminY}\, ,
\eeqn
$\cpY$ can be obtained by simply exchanging $X \getsto Y$, and $s_{\rm
  n/p} \equiv \Srmn/\Srmp$.  However, in order to reduce the
statistical uncertainties contributed from estimate of $\armn / \armp$
involved in $\calCpX$ and $\calCpY$, one can use one target with the
SD sensitivity (almost) only to protons or to neutrons combined with
another one with only the SI sensitivity.  For this case $\calCpX$ is
independent of $\armn / \armp$ and the expression
(\ref{eqn:rsigmaSDpSI}) for $\sigmapSD / \sigmapSI$ can be reduced to
\beq \label{eqn:rsigmaSDpSI_p}
   \frac{\sigmapSD}{\sigmapSI}
% = \frac{\FSIQthreY (\calR_{m, X}/\calR_{m, Y}) - \FSIQthreX}
%        {\calCpX \FSDQthreX}
 =-\frac{\FSIQthreX \calR_{m, Y} - \FSIQthreY \calR_{m, X}}
        {\calCpX \FSDQthreX \calR_{m, Y}}\, .
\eeq

\begin{figure}[t!]
\includegraphics[width=0.45\textwidth]{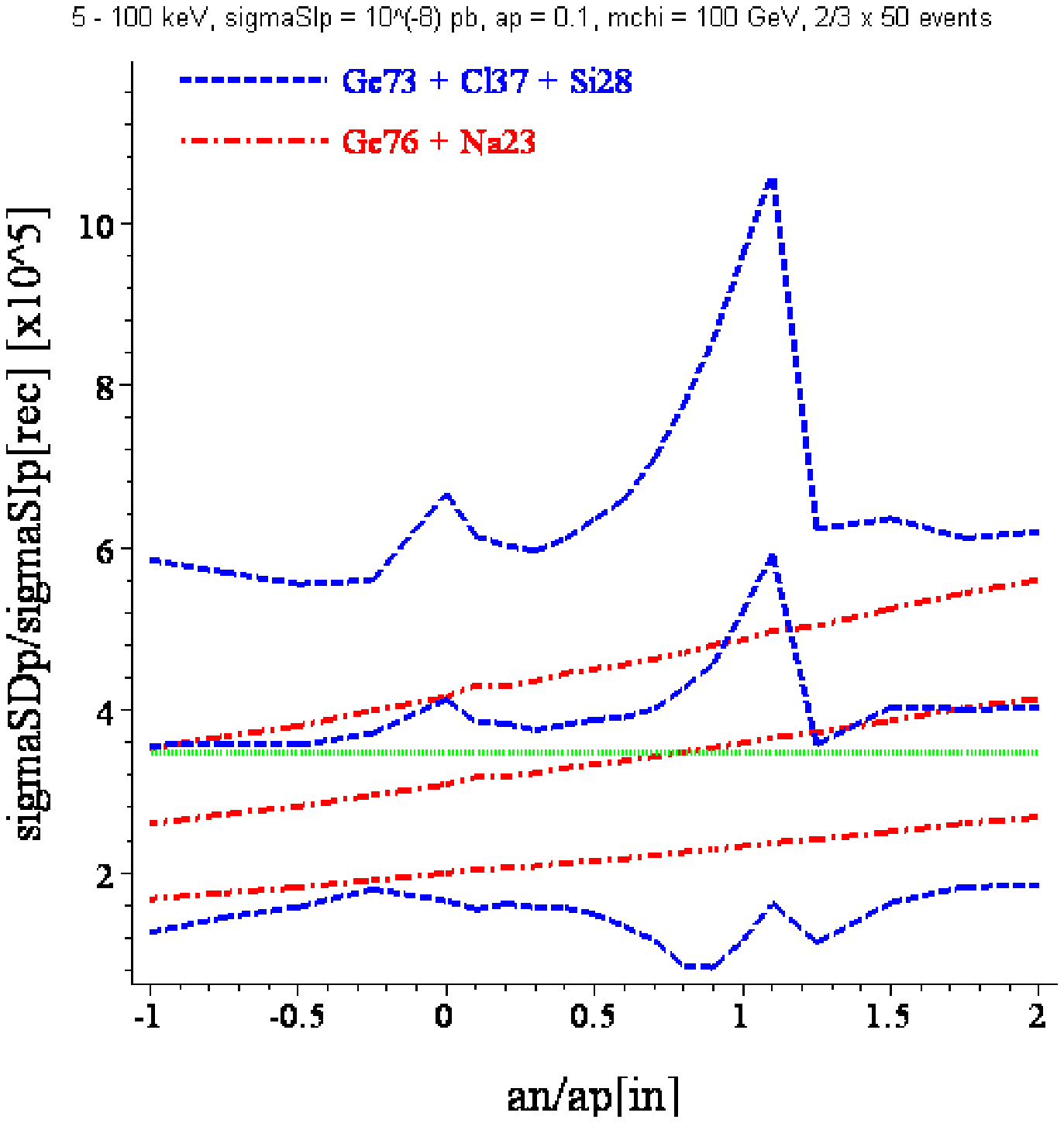} \hspace{0.025\textwidth}
\includegraphics[width=0.45\textwidth]{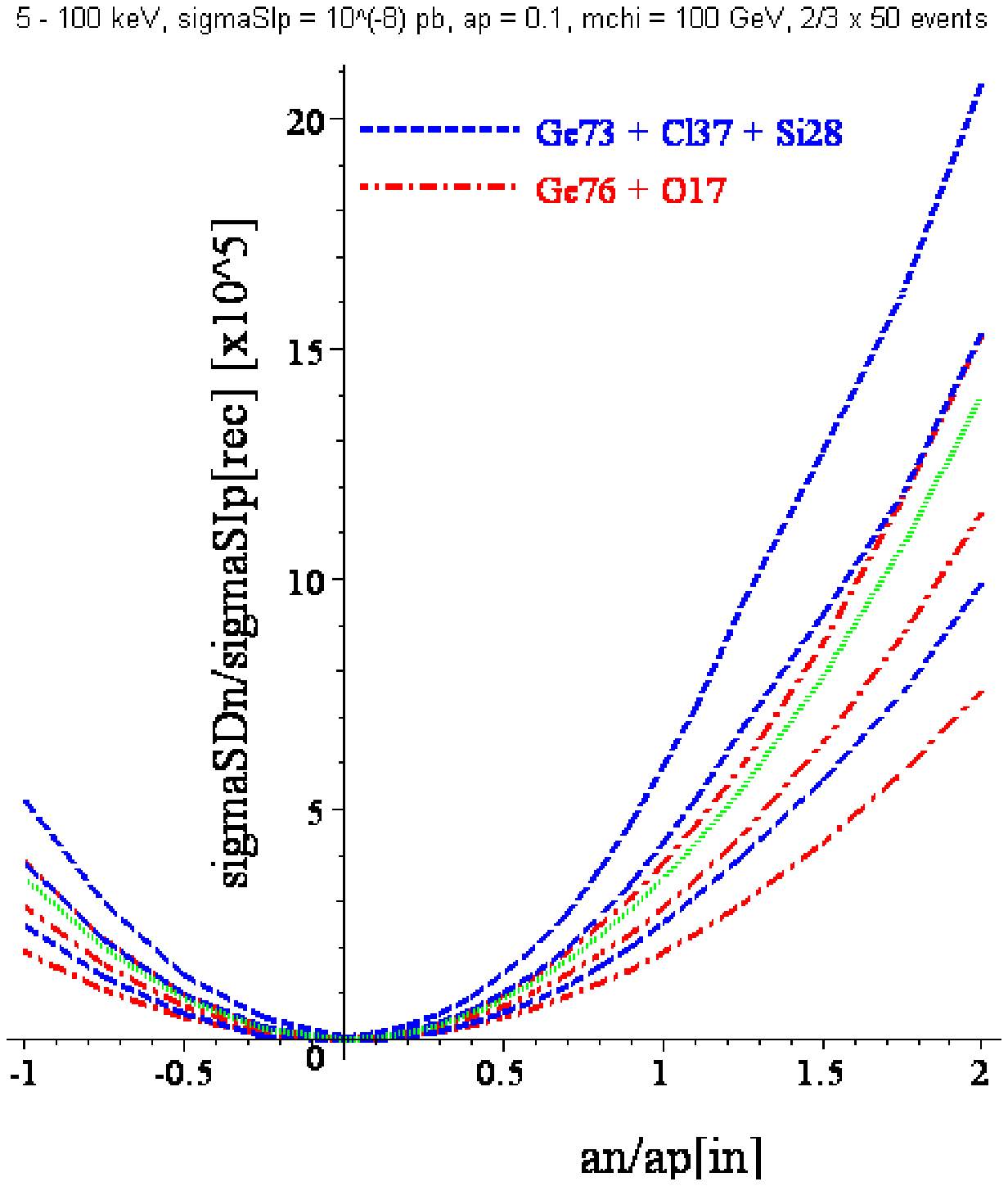}
\caption{Preliminary results for the reconstructed $\sigmapSD /
  \sigmapSI$ (left) and $\sigmanSD / \sigmapSI$ (right) as functions
  of the true (input) $\armn / \armp$, respectively.  The dashed
  (blue) curves indicate the values estimated by
  Eq.(\ref{eqn:rsigmaSDpSI}) with $\armn / \armp$ estimated by
  Eq.(\ref{eqn:ranapSISD}); whereas the dash--dotted (red) curves
  indicate the values estimated by Eqs.(\ref{eqn:rsigmaSDpSI_p}).
  $\sigmapSI$ and $\armp$ have been set as $10^{-8}$ pb and 0.1,
  respectively.  The other parameters are as in
  Figs.~\ref{fig:ranap_SD}. Note that, since we fix $\sigmapSI$ and
  $\armp$, $\sigmapSD / \sigmapSI$ shown here is a constant, whereas
  $\sigmanSD / \sigmapSI \propto \armn^2$ a parabola.}
\label{fig:rsigmaSDSI}
\end{figure}
Figures \ref{fig:rsigmaSDSI} show the reconstructed $\sigmapSD /
\sigmapSI$ (left) and $\sigmanSD/\sigmapSI$ (right) estimated by
Eqs.(\ref{eqn:rsigmaSDpSI}) and (\ref{eqn:rsigmaSDpSI_p}) as functions
of the true (input) $\armn / \armp$, respectively.  Besides
$\rmXA{Ge}{73}$ and $\rmXA{Cl}{37}$, $\rmXA{Si}{28}$ has been chosen
as the third target for estimating $\armn / \armp$ by
Eq.(\ref{eqn:ranapSISD}); whereas $\rmXA{Ge}{76}$ has been chosen as
the second target having only the SI interaction with WIMPs and
combined with $\rmXA{Na}{23}$ (for $\sigmapSD / \sigmapSI$) and
$\rmXA{O}{17}$ (for $\sigmanSD / \sigmapSI$) for using
Eq.(\ref{eqn:rsigmaSDpSI_p}).  We see here that, since the SD
WIMP--nucleus interaction doesn't dominate for our simulation setup,
$\sigmapSD / \sigmapSI$ estimated by Eq.(\ref{eqn:rsigmaSDpSI}) has
two discontinuities around the intersections at $\armn / \armp = -0.079 $ and
especially at $\armn / \armp = 1.16$, the intersection determined by
the $-\Srmp / \Srmn$ value of $\rmXA{Cl}{37}$.  However, from two
experiments with only $\sim$ 20 events in the low energy range, one
could in principle already estimate $\sigmapSD / \sigmapSI$ and
$\sigmanSD / \sigmapSI$ by using Eq.(\ref{eqn:rsigmaSDpSI_p}) with a
statistical uncertainties of $\sim$ 35\%.
\section{Summary and conclusions}
In this article we described model--independent methods for
determining the WIMP mass and their couplings on nucleons by using
future experimental data from direct Dark Matter detection.  The main
focus is {\em how well} we could extract the nature of WIMPs with {\em
  positive} signals and {\em which problems} we could meet by applying
these methods to (real) data analysis.

In Secs.~2 and 3 we discussed the determinations of the WIMP mass and
its spin--independent coupling on protons.  If WIMPs are light ($\mchi
\simeq$ 50 GeV), with $\cal O$(50) events from one experiment, their
mass and SI coupling could be estimated with errors of $\sim$ 35\% and
$\sim$ 15\%, respectively.  However, in case WIMPs are heavy ($\mchi
\gsim$ 200 GeV), the statistical fluctuation by the algorithmic
procedure for matching the maximal cut--off energies of the
experiments could be problematic for estimating their mass, and
thereby their SI coupling.

In Sec.~4 we turned to consider the spin--dependent interaction. The
simulations show pretty small statistical uncertainties.
 Moreover,
 differing from the traditional method
 for constraining the SD WIMP--nucleon couplings%
 \cite{{Tovey00}, {Giuliani04}, {Giuliani05}, {Girard05}},
 we do not make any assumptions
 on $\rho_0$, $f_1(v)$, and $\mchi$.
 The price one has to pay for this is that
 positive signals in at least two different data sets
 with different target nuclei are required.
%
% We extract the required information
% on the latter two quantities,
% which forces us to assume the existence
% of an additional data set in the analysis.
%
 In addition,
 without independent knowledge of $\rho_0$,
 one can only determine ratios of cross sections.
% or the product $\rho_0 \sigma$.

In summary, once two (or more) experiments measure WIMP events, the
methods presented here could in principle help us to extract the
nature of halo WIMPs.  This information will allow us not only to
constrain the parameter space in different extensions of the Standard
Model, but also to confirm or exclude some candidates for WIMP Dark
Matter%
\cite{{Bertone07}, {Barger08}}.
\section*{Acknowledgments}
This work was partially supported by the Marie Curie Training Research
Network ``UniverseNet'' under contract no.~MRTN-CT-2006-035863, by the
European Network of Theoretical Astroparticle Physics ENTApP ILIAS/N6
under contract no.~RII3-CT-2004-506222, as well as by the BK21
Frontier Physics Research Division under project no.~BA06A1102 of
Korea Research Foundation.
\end{document}